\documentstyle[twocolumn,prl,aps]{revtex}
\begin{document} 
\addtolength{\textheight}{8mm}
\title{
Matrix Elements 
between Nuclear Compound States and Dynamical Enhancement\\
of the Weak Interaction
}
\author{V. V. Flambaum and O. K. Vorov}
\address{$^*$ Theoretical Physics Department, School of Physics,
University of New South Wales, Sydney,
NSW 2052, Australia
\\
(Received  9 December 1992)
}
\address{\mbox{ }}
\address{\mbox{ }}
\address{\parbox{14cm}{\rm \mbox{ }\mbox{ }
A method to calculate the mean squared matrix element of weak
interaction between compound states is developed. 
The result is expressed 
in terms of matrix elements of the nucleon-nucleon strong 
and weak interactions
times the Fermi distribution functions at finite temperature.
Numerical calculations for $^{233}$Th are in excellent agreement 
with recent measurements of parity nonconservation effects in neutron
capture.
In fact, our calculations prove that the factor
of dynamical enhancement
(ratio of compound-nucleus effect to single-particle one) really exceeds
100, thus making it unnesessary to assume a value of the weak constant
bigger than 
standard one ($g\simeq 10^{8} \epsilon \sim 1\div 4$).      
}}
\address{\mbox{ }}
\address{\parbox{14cm}{\rm PACS numbers: 24.60.Dr, 24.60.Lz, 25.40.Dn,
24.80.Dc
}}
\maketitle
\makeatletter
\global\@specialpagefalse
\def\@oddhead{REV\TeX{} 3.0
\hspace{1.0cm}
Published in Phys.Rev.Lett,
\hspace{0.8cm} V. 70,  p.4051 
\hspace{0.8cm} 28 JUNE 1993}
\let\@evenhead\@oddhead
\makeatother
The parity nonconcerving (PNC) nucleon interaction 
in nuclei now attracts
attention of both experimentalists and theorists, especially,
in connection with the recent experiments on slow 
neutron scattering through
the compound nuclear states, where the measured PNC effect
(dependence of cross section on neutron helicity) 
proved to be of order several per cent \cite{1},\cite{2}
(cf. with the PNC effects in p-p or p-$\alpha$ scattering 
where effect is
of order of $3 \cdot 10^{-7}$).
Moreover, correlation of sign
in the effect on close neutron resonances has been observed \cite{2}.

In the current literature 
\cite{JBY,UV,lar},
two different approaches to explain 
a great value of this effect coexist, based on different assumptions and
contradicting each other.    The first one, the
statistical model of dynamical enhancement,
was considered in works \cite{Haas},\cite{SF}.(In fact, a large value
of the effect was predicted in the 1980-1981 papers of Ref.\cite{SF}.)
Within this approach, the essential
enhancement of the parity violating amplitude in neutron capture
arises from the mixing of closely lying (within interval of several eV)
compound nuclear
levels of opposite parity, and statistical enhancement $(\sim\sqrt{N})$
of the weak matrix element between compound wave functions composed of
$N\sim3\times10^{5}$ many-particle configurations. The
estimate of the magnitude of the effect  was given in \cite{SF},
\cite{JBY},\cite{UV},
based on the standard 
Hamiltonian of the weak interaction of nucleons in a nucleus
\begin{equation}
W=\frac{Gg}{2\sqrt{2}m}\lbrace({\bf \sigma}{\bf p}),\rho\rbrace,
\qquad \qquad \varepsilon=1.0 \cdot 10^{-8}g, 
\end{equation}
where $G=10^{-5}m^{-2}$ is the Fermi constant, $m$ is the nucleon mass, 
${\bf p}$ and ${\sigma}$ are the neutron momentum and its doubled spin 
correspondingly, while $\rho$ is the nuclear density; the nucleon
dimensionless constant $g_{p,n}$ (see e.g. Ref. \cite{DDH}) 
is of order unity
(now the  notation $\varepsilon$ is widely used \cite{lar}).

The second approach, the so called ``valence mechanism'' \cite{ZS}
(see also \cite{ND}), is based on
the assumption that the weak amplitude admixing the s-wave
to initial p-wave is dominated by direct matrix elements of PNC-potential
(1) between these states, so the effect is 
assumed to be of single-particle nature.
The valence mechanism gives correlations of the sign in the effect
on the different resonances but 
to explain the observed magnitude of the
effect in this approach, one has to use the neutron constant 
in Eq.(1) being at least two orders of magnitude larger than is predicted
by theory (see Refs.\cite{ND},\cite{lar}). 
Thus, the two mechanisms require a neutron 
weak interaction constant which differs by two orders of magnitude in the
range 1-300. It becomes even more important in connection with the
so-called ``Tsinoev puzzle'' \cite{Tsin} where the observed PNC effect
is $10^{3}$
times bigger than the theoretical estimate.

Staying within the framework of the first approach (dynamical enhancement) 
we present
here a method to calculate the mean squared weak matrix element between
s and p compound resonances, in the statistical 
model with account for nuclear
structure and realistic residual nucleon interaction. 
We should note that the matrix element between compound states was considered
earlier in Refs. \cite{JBY},\cite{UV}. However, these works
use some hypotheses which are not easy to justify. In the work \cite{JBY}
a proportionality relation between the matrix elements of the weak and
residual interaction was used. In the work \cite{UV} it was supposed that
the matrix element between compound states is given by the same formula
as the matrix element of the nucleon excitation from the ground state with 
only some
minor modifications (occupation numbers and the optical potential 
depend on the temperature of the compound nucleus). In our approach we have
not used these hypotheses, and our result looks different (e.g., it is not
proportional to the square of the frequency of the time-dependent field
$\omega^{2}$ or $T^{2}$ ($\omega = 0$ for the weak interaction, and $T$ is
a temperature)).

Note, that we do not discuss sign correlations in the present paper.

Calculation of mean squared weak matrix element    
is based on the equivalence theorem of canonical and microcanonical 
ensembles for a system with a large number of degrees of freedom. Let us 
remember that the wave function of any compound state with angular
momentum $j$ and parity $\pi$ may be expressed as follows
\begin{equation}
|j^{\pi})=\sum_{\alpha} C_{\alpha}|\alpha>, \qquad
|\alpha>=(a^+bc^+de^+ ...)_{j^{\pi}}|0>,
\end{equation}
through their components $|\alpha>$  being many-particle excitations over
the shell-model
ground state $|0>$ (we will denote them by simple Dirac
brackets saving the notation $| )$ for compound wave functions; here and in
what follows notation $(...)_{j\pi}$
means the coupling of nucleon creators $a^{+}$ and destructors $a$
to total angular momentum $j$ and
parity $\pi$). Amongst them, it is reasonable to separate explicitly
the contribution of the ``principal components'', $|j^{\pi}))$,
dominating normalisation of the compound state, Eq.(2). The energies of these
components must be within the interval 
$[E-\frac{\Gamma_{spr}}{2},E+\frac{\Gamma_{spr}}{2}]$, 
where $E$ is the energy of the compound
state and $\Gamma_{spr}$ is the spreading width of the 
component (typically, $\Gamma_{spr} \sim 2MeV$, Refs.\cite{BM,SF}).
These components
(which contain several excited nucleons) can be composed by excitations
of protons and neutrons only inside
incomplete (valence) shells.
Mean squared values of the coefficients $C(E_{\alpha}$ 
can be described by the formulae (see e.g., Ref.\cite{BM})
\begin{eqnarray}
\overline{C^{2}(E_{\alpha})}=\frac{1}{\overline{N}}\Delta(\Gamma_{spr},
E-E_{\alpha}),\qquad \overline{N}=\frac{\pi\Gamma_{spr}}{2d},\\ \nonumber 
\Delta(\Gamma_{spr},E-E_{\alpha})=
\frac{\Gamma_{spr}^{2}/4}{(E-E_{\alpha})^{2}+\Gamma_{spr}^{2}/4},
\end{eqnarray}
where $E_{\alpha}$ is the energy of an arbitrary many-particle configuration),
$d$ is the averaged energy distance between the resonances, and $\overline{N}$
is the number of principal components.
The Breit-Wigner-type factor $\Delta$, describing cutting off of weights 
before states distanced in energy, may be treated as a ``spread'' 
$\delta$-function normalized as to be of
order unity for $|E-E_{\alpha}|\leq\Gamma_{spr}/2$ and with conventional
limit $\Delta(\Gamma_{spr},E-E_{\alpha})
\to$ $\frac{\pi\Gamma_{spr}}{2}\delta(E-E_{\alpha})$ for $\Gamma_{spr}\to 0$.
Thus, the
coefficients before the ``principal'' components $\tilde C_{\alpha}$
in (2) are governed by the microcanonical ensemble rule \cite{BM,SF}.
The very important point should be born in mind, that  there 
are no single-particle states of opposite parity having the 
same angular momenta within the valence shells. Since the weak interaction
(Eq.(1)) mixes only states of such type, 
it follows from the definition of ``principal'' components that the
weak matrix element between two compound state of close energy is
dominated by the weak transitions between ``small'' components
of one resonance and ``principal'' ones of the second resonance, and
vice versa (this was firstly mentioned by the authors of Ref.\cite{ZS}, see
also \cite{SF}).
Any transfer of one particle from the valence shell to another one gives a
rise in excitation energy as large as $E_{sp}\sim 8 MeV$ (what is much more 
than typical matrics elements of the residual interaction $V$)
leading out a 
configuration from the microcanonical ensemble of ``principal'' components
according
to Eq.(3). Therefore, one can easily generate the appropriate set of ``small''
configurations by means of first-order perturbation theory in
the residual strong interaction $V$. Thus, matrix elements of the weak 
interaction between compound states look like 
\begin{eqnarray}
(s|W|p)\quad = \quad
\sum_{\alpha}\frac{((s|V|\alpha><\alpha|W|p))}{E-E_{\alpha}}+
\nonumber\\
\sum_{\beta}\frac{((s|W|\beta><\beta|V|p))}{E-E_{\beta}},
\end{eqnarray}
where $|\alpha>$ and $|\beta>$ are small components, and compound states
$|s))$, $|p))$ contain only principal components. 
We stress that we do not need any ``exotic'' parts
of the residual strong interaction
here. Since $E-E_{\alpha} \sim 8MeV \gg V$ (see above), only the dominating 
and well-known parts \cite{Landau,Migdal,BG} of the two-nucleon interaction
\begin{equation}
V \quad = \quad \frac{1}{2}\sum_{ab,cd}a^{+}bV_{ab,cd}c^{+}d, 
\end{equation}
which will be specifyied below, are important in Eqs.(4),(5). Here, our
consideration is general and not even confined to nuclear system.
 
The weak interaction (Eq.(1)) is a single-particle operator.
This fact allows one to the include weak interaction into
the mean nuclear field
and transfer the perturbation theory expansion in the single-particle orbitals:
$\tilde{\psi}_{a}=\psi_{a}+
\sum_{A}\frac{<\psi_{A}|W|\psi_{a}>}{\epsilon_{a}-\epsilon_{A}}\psi_{A}$,
where $\epsilon_{a}$ and $\epsilon_{A}$ are the energies of the orbitals
$\psi_{a}$ and $\psi_{A}$ (differing by their parities), 
the large Latin indices label the corresponding off-valence-shell states.
Thus, we can express result in terms of the residual interaction renormalised
by the weak interaction
(${\tilde V}_{ab,cd} =
V(\tilde{a} \tilde{b}, \tilde{c} \tilde{d})$): 
\begin{eqnarray}
{\tilde V}_{abcd} =
\sum_{A}\frac{V_{Ab,cd}}{\epsilon_{a}-\epsilon_{A}}w_{aA}
+ \sum_{B}\frac {V_{aB,cd}}{\epsilon_{b}-\epsilon_{B}}w_{Bb} +
\nonumber\\
\sum_{C}\frac{V_{ab,Cd}}{\epsilon_{c}-\epsilon_{C}}w_{cC} + 
\sum_{D}\frac{V_{ab,cD}}{\epsilon_{d}-\epsilon_{D}}w_{Dd},
\end{eqnarray}
here $w_{aA} \equiv <\psi_{A}|W|\psi_{a}>$. With these notations, 
to the first order in $V$, Eq.(4)
can be read as follows:
$(s|W|p) = ((s|\tilde{V}|p))$.
The advantage of using the effective two-particle PNC-interaction
$\tilde V$ is that the matrix elements between compound states 
are expressed through the matrix elements ${\tilde V}_{ab,cd}$ between
valence shell
single-particle states. Thus we avoid the necessity of explicitly considering
the ``small'' components of the compound states which we believe cannot  be
described by the same spreading widths as the principal
components (see Eq.(3)).

Consider now the mean squared value of this matrix element:
\begin{eqnarray}
\overline{(p|W|s)(s|W|p)} \quad =
\quad \overline{((p|\tilde{V}|s))((s|\tilde{V}|p))}
= 
%
\nonumber\\
=\sum_{\alpha \beta} \overline{C_{\alpha}C_{\beta} ((p|\tilde V|\alpha>
<\beta|\tilde V|p))} \quad =
\nonumber\\
=\quad \sum_{\alpha} \frac{1}{\overline{N}}
\Delta(\Gamma_{spr},E-E_{\alpha})
\overline{((p|\tilde{V}|\alpha><\alpha|\tilde{V}|p))}.
\end{eqnarray} 
Here, we have expanded the compound state $|s)$ in terms
of the components (2) and
made use of the statistical independence of the coefficients $C_{\alpha}$ 
(see Eqs.(2,3),Ref.\cite{BM},\cite{lar},\cite{Haas},\cite{SF}): 
\begin{equation}
\overline{C_{\alpha}C_{\beta}}\quad =\quad \overline{C^{2}_{\alpha}}
\delta_{\alpha \beta}\quad=
\quad \delta_{\alpha \beta} \frac{1}{\overline{N}}
\Delta(\Gamma_{spr},E-E_{\alpha}).
\end{equation} 
In the second quantization representation, summation over $\alpha$ in (7)
is equivalent to summation over different components of operator the $V$
in Eq.(5), i.e. the problem is reduced
to the calculation of $((p|\tilde{V} \tilde{V}|p))$.
Then, to calculate the averaging over p-resonance ``principal'' components
$\overline{((p| ... |p))}$ in $\overline{W^{2}}$, let us use, instead
of the present microcanonical ensemble, the equivalent canonical one
(which may be choosen for a system with a large number degrees of freedom by
introducing the effective nuclear temperature T and chemical potentials 
$\lambda_{n},\lambda_{p}$). In such a way, the expectation value
in (8) is reduced to a
canonical ensemble average with the standard contractor rules
$((p|\overline{a^{+}b}|p))=\delta_{ab}\nu^{T}_{a}$, for
$\nu^{T}_{a}$ being the finite temperature Fermi occupation probabilities,
$\nu^{T}_{a}=\lbrace exp[(\epsilon_{a}-\lambda)/T]+1
\rbrace ^{-1}$. The canonical ensemble parameters $T$, $\lambda_{\tau}$ ($\tau$
means isospin projection) are to be determined from 
conventional ``consistency'' equations $E = \sum_{a}\nu_{a}\epsilon_{a}$, 
$ Z=\sum_{p}\nu_{p}$, and $N=\sum_{n}\nu_{n}$ 
for the excitation energy $E$ (being equal the to neutron separation energy,
$B_{N}$),
nuclear charge $Z$, and neutron number $N$ correspondingly.     
After contractor evaluations one simply obtains from (7):
\begin{eqnarray}
\sqrt{\overline{W^{2}}}= 
\sqrt{\frac{2d}{\pi\Gamma_{spr}}}
\times \qquad  \qquad \qquad \qquad \qquad \qquad \\ \nonumber
\times \biggl\{ \sum_{abcd}\nu^{T}_{a}(1-\nu^{T}_{b})
\nu^{T}_{c}(1-\nu^{T}_{d})\frac{1}{4}
 \mid \tilde V_{ab,cd} - \tilde V_{ad,cb} \mid ^{2}
\nonumber\\
\Delta(\Gamma_{spr},
\epsilon_{a}-\epsilon_{b}+\epsilon_{c}-\epsilon_{d})\biggr\} ^{\frac{1}{2}}.
\end{eqnarray}
The argument of the function $\Delta$ here is the change of the
energy: $E-E_{\alpha}=\epsilon_{a}-\epsilon_{b}+\epsilon_{c}-\epsilon_{d}$,
and $\tilde{V}$ is given by Eq.(6). In fact, it is an approximate energy 
conservation law with the accuracy up to width of states.

The numerical calculations for $^{233}$Th have been performed with the use 
of single-particle basis of states obtained by numerical calculations
of the eigenvalue
problem for the Woods-Saxon potential with spin-orbital interaction  in the
form 
$U(r)=-U_{0}f(r)+U_{ls}({\bf \sigma l})(\hbar/(m_{\pi}c))^{2}\frac{1}{r}
\frac{df}{dr}+ U_{c}$ with $f(r)=(1+exp((r-R)/a))^{-1}$, where
${\bf l}$ is the orbital angular momentum, $U_{c}$ means
Coulomb correction for protons, $U_{c}=3Ze^{2}/(2R)(1-r^{2}/(3R^2)), r\leq R$ 
and $U_{c}=Ze^{2}/r, r>R$, for $R$, $a$, and $r$ being the nuclear radius, 
diffusity parameter and radial variable correspondingly. The parameter values 
were used in accordance with Bohr-Mottelson formulae (see Ref.\cite{BM})
for the 
case of $^{233}Th$: they are close to those established for heavy nuclei like
lead (Ref.\cite{BG}) to reproduce single-particle properties.
As for the residual 
interaction, we have employed the most widely used Landau-Migdal particle-hole 
interaction of contact type with spin- and isospin-exchange terms which rises
to Landau Fermi liquid theory (Ref.\cite{Landau}); for the case of a nucleus it
was established in the Theory of Finite Fermi Systems \cite{Migdal,BG} by
summation of all
graphs irreducible in the particle-hole direction. In that case, the
explicit form of the matrix $V$ in (5) is given by the second
quantized version of the interaction 
\begin{equation}
V({\bf r}_,
{\bf r'})=C\delta({\bf r}-{\bf r}')[f+f'{\bf \tau}{\bf \tau'}+
g{\bf \sigma}{\bf \sigma'}+g'{\bf \tau}{\bf \tau'}{\bf \sigma}{\bf \sigma'}],
\end{equation}
where $C=300$ $MeV$ $fm^{3}$ is the universal
 Migdal constant \cite{Migdal},\cite{BG}, ${\bf \sigma}({\bf \sigma}')$
and ${\bf \tau}({\bf \tau}')$ mean particle(hole) spin and isospin Pauli
matrices, respectively, and the 
strenghs $f,f',g,g'$ are in fact functions of $r$ via density dependence:
$f=f_{in}-(f_{ex}-f_{in})(\rho(r)-\rho(0))/\rho(0)$ (the same for $g,g'$) with
ansatz $\rho(r)=\rho(0)f(r)$ (see above). 
(Quantities subscripted by ``in'' and ``ex''
characterize interaction strenghs in the depth of the nucleus and on 
its surface, respectively). This interaction, with its parameter values
listed below, has been successfully used by many authors
(see Refs. \cite{BG}) to quantitatively describe a great amount of 
various properties of heavy nuclei. 
The value of temperature $T=0.6MeV$ was used in accordance to the consistency
condition for excitation energy (see above). 

We present here the results for the conventional choice of values
for parameters in the Landau-Migdal interaction, Eq.(10), which has been
widely used for
heavy nuclei  
(see \cite{Migdal},\cite{BG} and references therein), 
namely $f_{ex}=-1.95$, $f_{in}=-0.075$ $f'_{ex}=0.05$ $f'_{in}=0.675$,
$g_{in}=g_{ex}=0.575$, and $g'_{in}=g'_{ex}=0.725$.   
Note that the exchange matrix elements of $V$ enter in Eq.(9) 
but, generally, the values of the parameters
$f$, $f'$, $g$, and $g'$ are chosen in such a way that the exchange terms
are already included (it can always be done by use of the Firtz 
transformation).
The variant referred to below as I corresponds to this standard procedure;
we present also the results for the case when the excange terms in (9)
are included explicitly (referred to as II).   
The results for $\sqrt{\overline{W^{2}}}$ (8) may be
expressed explicitly in terms of 
the proton and neutron weak constants $g_{p}$ and $g_{n}$ in the form: 
\begin{equation}
\sqrt{\overline{W^{2}}}=\frac{1}{\overline{N}}\sqrt{\overline{N}}
\sqrt{(\Sigma_{pp}g_{p})^{2}
+(\Sigma_{nn}g_{n})^{2}+\Sigma_{pn}g_{p}g_{n}}(meV);
\end{equation}
the numerical calculations of the constants $\Sigma$ gives
the following results:
\begin{displaymath}
I: \qquad \sqrt{\overline{W^{2}}}= 2.08 meV ;
\qquad \quad II: \qquad \sqrt{\overline{W^{2}}}= 3.57 meV .     \nonumber\\  
\end{displaymath}
The experimental value is $\sqrt{\overline{W^{2}}}=1.39^{+0.55}_{-0.38}meV$
(\cite{2}).
The values of $\sqrt{\overline{W^{2}}}$ for the cases I and II were
obtained for the conventional choice of weak constants,
$g_{p}=4$ and $g_{n}=1$ (Ref.\cite{ND}).
(In the notation widely adopted in the current 
literature \cite{2}, for the neutron case it corresponds to the value
$\varepsilon=10^{-8}g_{n}=10^{-8}$).
Now, we can compare this result with single-particle (valence) estimation
$w_{val}$. In the valence mechanism, only single-particle components contribute
(in $^{233}$Th, $4s$
and $4p$ neutron states). Therefore, the valence model result is
\begin{equation}
w_{val}\simeq \frac{1}{\overline{N}}<4s|W|4p>\simeq 
\frac{1}{\overline{N}} g_{n}0.740(eV) = 1.72 \cdot 10^{-3}meV.
\end{equation}
Thus, statistical 
contribution is $10^{3}$ times bigger (due to the extra factor
$\sqrt{\overline{N}}$, compare Eq.(9,11) with (12)). This factor reflects
the incoherent contribution of all $N$ components. 
Calculations fulfilled for other sets of parameters
display no strong sensitivity of the numerical results to variations
of both single-particle basis and residual interaction.
Let us point out that in our approach,
the only essential assumption made is that of the statistical properties of 
the distribution for coefficients $C_{\alpha}$ in Eqs.(2),(8). As 
uncertainity in definition of $\overline{N}^{-1/2}$ is concerned, two estimates
for it, namely $\overline{N}^{-1/2} \simeq \sqrt{\frac{2d}{\pi \Gamma_{spr}}}$
(for $d=17 eV$ in $^{233}$Th case and spreading $\Gamma_{spr} \simeq 2MeV$)
and that from the widths of the compound resonances,
$\overline{N}^{-1/2} \simeq \sqrt{<\Gamma_{n}/\Gamma^{0}_{n}>}$ give
approximately
equal values $\overline{N}^{-1/2} \simeq 2.3 \cdot 10^{-3}$ with accuracy up 
to a factor of 2 (here $\Gamma^{0}_{n}$ is the width of the s or p
single-particle
resonance (see Ref.\cite{BM}), and $\Gamma_{n}$ is the width of the $s$ or $p$ 
compound resonance correspondingly).

The results of this work can be summarised as follows. A consistent
method to calculate the mean squared weak matrix element between two compound 
states of opposite parity is proposed, based on a statistical model with
account for nuclear structure. The results prove
that the dynamical enhancement $\sim \sqrt{\overline{N}}$
does really exist. The observed large value of
this quantity for $^{233}Th$ is explained in terms of the model with the
conventional neutron weak constant. Moreover, even for the zeroth value
of the neutron constant the effect is still reproduced without substantial 
cut-off because of the proton PNC transitions in the nuclear compound state.
(In fact, even for $g_{n}=1$, proton contribution is few times larger
 due to the bigger value of constant $g_{p}$).
Further experiments in this region would be of great importance.

We are grateful to V.F.~Dmitriev and V.B.~Telitsin for kindly providing us
with a computer program to solve eigenvalue problem in Saxon-Woods potential.

\noindent

\begin{thebibliography}{200}
\bibitem{1} V.P.~Alfimenkov {\it et al}, Pis'ma Zh. Eksp. Theor. Fiz. {\bf 34}
, 308 (1981) [JETP Lett. {\bf 34}, 295 (1981)]; Nucl. Phys. {\bf A398}, 93
(1983); Usp. Fiz. nauk {\bf 144}, 361 (1984) [Sov. Phys. Usp. {\bf 27}, 797
(1984)].
\bibitem{2} J.D.~Bowman, C.D.Bowman {\it et al.}, Phys. Rev. Lett. {\bf 65},
1192 (1990),C.M.~Frankle {\it et al.}, Rev. Lett. {\bf 67}, 564 (1991).
\bibitem{JBY}
M.B.Johnson, J.D.Bowman, and S.H.Yoo, Phys.Rev.Lett. {\bf 67}, 310 (1991).
\bibitem{UV}
M.H.~Urin and O.N.~Vyazankin, Phys.Lett. {\bf B269}, 13 (1991).
\bibitem{lar} 
H.A.~Weidenm\"uller, Nucl.Phys. {\bf A522}, 293c (1991);  V.V.Flambaum, Phys.Rev. {\bf C45}, 437 (1992); J.D.~Bowman, G.T.Garvey, C.R.Gould, A.C.Hayes, and M.B.Johnson, Phys.Rev.Lett., {\bf 68}, 780 (1992);
N.Auerbach, Phys.Rev. {\bf C45}, R514 (1992); S.E.Koonin, C.W.Johnson, and
P.Vogel, Phys.Rev.Lett. {\bf 69}, 1163 (1992); N.Auerbach and J.D.Bowman,
Preprint MSUCL-830, 1992.
\bibitem{Haas} R.~Haas, L.B.~Leipuner, and R.K.~Adair, Phys. Rev. {\bf 116},
1221 (1959); R.J.~Blin-Stoyle, Phys. Rev. {\bf 120}, 181 (1960); I.S.~Shapiro
, Usp. Fiz. Nauk {\bf 95}, 647 (1968) [Sov. Phys. Usp. {\bf 11}, 682 (1969)].
\bibitem{SF} O.P.~Sushkov and V.V.~Flambaum, Pis'ma Zh. Eksp. Theor. fiz.
{\bf 32}, 377 (1980) [JETP Lett. {\bf 32}, 353 (1980)]; INP Reports Nos. 
80-148, 1980 and 81-37, 1981; Usp. Fiz. Nauk {\bf 136}, 3 (1982) [Sov. Phys. 
Usp.
{\bf 25}, 1 (1982)]; 16th LINP Winter School Proceedings, p. 200, (Leningrad,
1981); Nucl. Phys. {\bf A412}, 13 (1984);
S.G.Kadmensky, V.P.~Markushev, and V.I.~Furman, Yad. Fiz. {\bf 37}, 581
(1983) [Sov. J. Nucl. Phys. {\bf 37}, 345 (1983)].
\bibitem{DDH} B.~Desplanques, J. Donoghue and B.~Holstein, Ann. of Phys.
{\bf 124},449 (1980); V.V.~Flambaum, I.B.~Khriplovich, and O.P.~Sushkov,
Phys. Lett. {\bf 146B}, 367(1984).
\bibitem{ZS} D.F.~Zaretsky and V.I.~Sirotkin, Yad. Fiz. {\bf 37}, 607 (1983) 
[Sov. J. Nucl. Phys. {\bf 37}, 361 (1983)]; {\bf 45}, 1302 (1987) [{\bf 45},
808 (1987)].
\bibitem{ND} S.~Noguera and B.~Desplanques, Nucl. Phys. {\bf A457},
189 (1986).
\bibitem{Tsin} L.V.~Inzhechik {\it et al.} Yad.Fiz. {\bf 44}, 1370 (1986)
[Sov.J.Nucl. Phys. {\bf 44}, 890 (1986)], Zh.Eksp.Theor.Fiz. {\bf 93},
800 (1987) [Sov.Phys.JETP {\bf 66}, 450 (1987)];
Zh.Eksp.Theor.Fiz. {\bf 93}, 1569 (1987) [Sov.Phys.JETP {\bf 66}, 897 (1987)];
O.P.~Sushkov and V.B.~Telitsin, submitted to Nucl. Phys. A.  
\bibitem{BM} A.~Bohr and B.~Mottelson, {\it Nuclear Structure} (Benjamin,
New York, 1969), Vols. 1 \& 2.  
\bibitem{Landau} L.D.Landau, Zh.Eksp.Theor.Fiz. {\bf 30}, 1058 (1956)
[Sov.Phys.JETP {\bf 30}, 920 (1956)]; Zh.Eksp.Theor.Fiz. {\bf 32}, 59 (1957)
[Sov.Phys.JETP {\bf 5}, 101 (1957)]; Zh.Eksp.Theor.Fiz. {\bf 35}, 97 (1958)
[Sov.Phys.JETP {\bf 8}, 70 (1959)]. 
\bibitem{Migdal} A.B.Migdal, {\it Theory of Finite Fermi Systems and 
Applications to atomic Nuclei} (John Wiley \& Sons, New York, 1967). 
\bibitem{BG} G.E.Brown, Rev.Mod.Phys., {\bf 43},1 1971);
V.Klemt, S.A.Moszkowski, and J.Speth, Phys.Rev. {\bf C14}, 302 (1976);
J.Speth, E.Werner, and W.Wild, Phys.Rep. {\bf 33}, No.3, 127(1977) and
references therein; 
G.Bertsch, D.Cha, and H.Toki, Phys.Rev. {\bf C24}, 533 (1981);
V.A.Khodel and E.E.Sapershtein, Phys.Rep. {\bf 92}, 183 (1982), and
references therein;
J.W.Negele, Rev.Mod.Phys. {\bf 54}, 913 (1982);
R.De Haro, S.Krewald, and Speth, Nucl.Phys., {\bf A388},
265 (1982); K. Goeke, and J.Speth, Annu. Rev. Nucl. Part. Sci. {\bf 32}, 65
(1982); F.~Osterfeld, Rev.Mod.Phys., {\bf 64}, 491 (1992), and references
therein. 
\end{thebibliography}
\end{document}